\documentstyle[preprint,aps]{revtex}

\begin{document} 
\draft
\title{Velocity Distributions \& Density Fluctuations in a 2D Granular Gas}
\author{J. S. Olafsen \cite{jso} and J. S. Urbach}
\address{Department of Physics,Georgetown University,Washington, D.C. 20057}
\date{\today}
\maketitle

\begin{abstract}
Velocity distributions in a vibrated granular monolayer are investigated
experimentally.  Non-Gaussian velocity distributions are observed at low
vibration amplitudes but cross over smoothly to Gaussian distributions
as the amplitude is increased.  Cross-correlations between 
fluctuations in density and temperature are present only when the 
velocity distributions are strongly non-Gaussian.  
Confining the expansion of the granular layer results in non-Gaussian 
velocity distributions that persist to high vibration amplitudes.

\end{abstract}


The effects of inelasticity on the statistical properties of a granular
gas has been a topic of recent intense theoretical and experimental interest.
In freely cooling granular media, analytic results and simulations show that
dissipative inter-particle collisions result in clustering \cite{McNamaraY96},
non-Gaussian velocity distributions \cite{EsipovP97}, and eventually to the
breakdown of hydrodynamics \cite{GoldhirschZ93}.  In driven granular gases,
where the energy lost through collisions is balanced by energy input from
external forcing, the effects of inelasticity can be observed in the 
steady state statistical properties of the gas.  Experimental studies
have shown clustering \cite{us,KudrolliWG97} and non-Gaussian 
velocity distributions \cite{us,LosertCDKG99,WarrJH94}, but the dynamical
origin of the velocity distributions remains unclear.  A model of a 
granular gas coupled to a thermal reservoir shows long-range correlations
in density and velocity, and velocity distributions that fall off with
$v^{3/2}$ in the tails \cite{NoijeE98}.  A model with a discrete random
forcing produces
strong clustering, and a cross-correlation between the fluctuations in 
density and granular temperature (average kinetic energy) \cite{puglisi}.
The cross-correlation is due to the same mechanism as the clustering
instability in a freely cooling granular gas:  fluctuations of increased
density result in more frequent inter-particle collisions, producing increased
dissipation and a reduced local granular temperature.

Recent work has demonstrated various collective phenomena for a large number
($\approx 20000$) of identical, uniform ball bearings constituting less than
one layer coverage on a vertically shaken, horizontal plate 
\cite{us,LosertCG98}.  At peak plate accelerations above 1 g and for most 
densities, the particles behave as a rapidly fluctuating gas. 
Reducing the acceleration amplitude decreases the mean square velocity,
or granular temperature, T$_G = <v^2>/2$, and effectively ``cools'' the gas,
leading to an observed increase in clustering. 
Continued cooling eventually leads to the formation of a collapse,  
a condensate of motionless particles that remain
in contact with the plate and each other.  In the range of 0.8 - 1.0 g, all
of the velocity distributions of the gas appear to scale with the second 
moment of the distribution to an universal curve \cite{us}.  
The velocity distributions demonstrate 
strong deviations from a Gaussian distribution in both the high velocity tails
and at low velocities.

This report presents the results of a further investigation into the nature
of the non-Gaussian velocity distributions and their relation to the observed
density fluctuations that stem from the inelastic collisions.
In order to understand the non-Gaussian velocity distributions seen previously
in this system \cite{us}, measurements were made
over a larger range of $\Gamma = A (2 \pi \nu)^2/g$, the dimensionless 
acceleration of the plate, where $\nu$ is the frequency in Hz and g is the 
acceleration due to gravity.  Our results demonstrate both 
non-Gaussian velocity distributions and a cross-correlation between density
and temperature, similar to that seen in the model system of Puglisi 
{\em et al.} \cite{puglisi}.  However, the data
reported here clearly shows that the two effects are independent:  there is
a large range over which the velocity distributions are non-Gaussian but for 
which the granular temperature is not measurably dependent on the density.  
Only at low $\Gamma$, where the clustering is the strongest, is the 
cross-correlation between density and temperature observed in our experiment.
  
For these experiments, the plate vibration is sinusoidal, $\nu=$ 70 Hz and 
unless otherwise stated, the particle density is $\rho = 0.532$. 
For accelerations below $\Gamma = 1.25$, the system was initially fluidized by 
shaking at $\Gamma = 1.25$.  
Two different particle species were used:  The small spheres were 302 
stainless steel with an average diameter of 0.1191 cm $\pm$ 0.00024 cm and
 the large spheres were 316 
stainless with an average diameter of 0.1588 cm $\pm$ 0.00032 cm.
The coefficient of restitution for both particle species is 
approximately 0.9 \cite{us}.

The measured velocity distributions at $\Gamma = 0.93$, $\Gamma=1.50$ and
$\Gamma=3.00$ are shown in Fig. \ref{fig:one}.  The distribution crosses 
over from
one with approximately exponential tails as reported in \cite{us} to a
Gaussian distribution.  This remarkable evolution is superficially similar
to what is observed in freely cooling granular media, where an initial
Gaussian velocity distribution becomes non-Gaussian as the system cools.
In that case, the evolution is determined by the strength of the 
inelasticity and the integrated number of collisions per particle, which does
not have an obvious analog in the driven system.

As previously reported, the non-Gaussian velocity distributions observed at 
low accelerations are accompanied by clustering, as demonstrated by a 
dramatic increase in the structure of the pair correlation function.  
Figure \ref{fig:two}(a) shows the pair correlation function determined via
analysis of images taken from above the granular layer for the same
accelerations as the velocity distributions in Fig. \ref{fig:one}.
The results of the measured correlation function approach that of an 
uncorrelated dilute hard sphere gas (solid line).  Thus, the crossover
to Gaussian velocity distributions is accompanied by the disappearance 
of spatial correlations, consistent with the suggestions that the 
non-Gaussian velocity distributions arise from a coupling between density
and temperature fluctuations \cite{puglisi}.

Increasing the steady state kinetic energy of the granular gas by increasing
the amplitude of the acceleration at constant frequency causes the gas to
change from primarily two-dimensional, where the particles never hop over
one another, to essentially three-dimensional \cite{movie}.  
This transition can be observed in the pair correlation function, $G(r)$, 
by the increase in its value for $r<1$. (The correlation function includes only
particle separations in the hortizontal plane.)
This transition can affect the dynamics in several
ways:  the effective density is decreased, so that excluded volume effects
are less important; the inter-particle collisions can occur at angles
closer to vertical, affecting transfer of energy and momentum from the 
vertical direction to the horizontal; and the change in the dimensionality
itself can have important consequences.  In order to separate these effects
from the direct consequences of increasing the kinetic energy of the gas,
a plexiglass lid was added to the system at a height of 0.254 cm, or 1.6
ball diameters for the larger particles.  For this plate-to-lid separation,
the larger particles cannot pass over top of one another, although 
enough room remains for collisions between particles at sufficiently different
heights to transfer momentum from the vertical to the horizontal direction.

Figure \ref{fig:two} (b) demonstrates the persistence of
the particle-particle correlations when the system is constrained 
to 2D.  The particle-particle correlation function decreases slightly from
$\Gamma = 0.93$ to $\Gamma = 1.50$, and then remains essentially 
constant up to $\Gamma = 3$.  The small value of $G(r)$ for distances less 
than one ball diameter indicates that the system remains 2D as $\Gamma$ 
is increased.  The structure observed in the correlation function is 
essentially the same as that of an equilibrium elastic hard sphere gas at 
the same density, indicating that the correlations that exist are due to 
excluded volume effects. 

Figure \ref{fig:three}(a) shows that the presence of the lid adds an energy
sink to the system at high $\Gamma$.  At low acceleration 
($\Gamma \leq 1$), very few, if any, particles strike the lid and the
lid has no significant effect.  At larger $\Gamma$ it is clear that 
the horizontal granular temperature is reduced
as a significant number of particles strike the 
lid, dissipating energy.  It is interesting to note that T$_G$ approaches
zero lineraly at finite $\Gamma$, indicating that the relationship between the
driving and the horizontal granular temperature is of the form 
T$_G \propto \Gamma - \Gamma_c$.

In order to demonstrate the effect of the lid on the velocity distributions,
we use a simple quantitative measure of the non-Gaussian nature, the flatness 
(or kurtosis) of the distribution:  

\begin{equation}
F = \frac{<v^4>}{<v^2>^2}.
\label{eq:f}
\end{equation}
\noindent
For a Gaussian distribution, the flatness 
is 3 and for the broader exponential distribution, the flatness is 6.  
In the absence of a lid, the flatness demonstrates
a smooth transition from non-Gaussian to Gaussian behavior as the granular 
temperature is raised (Fig. \ref{fig:three}(b)) whether the smaller
(circles) or larger (stars) particles are used.  
With the lid on, the velocity distributions remain more non-Gaussian
than in the free system for identical granular temperatures and 
density (diamonds).  The crossover from Gaussian to non-Gaussian behavior 
observed without the lid is therefore not simply an
effect of increasing the vertical kinetic energy of the particles, but 
rather related to the transfer of energy from the vertical to horizontal
motion in the system via collisions, the change in the density, or the
change in dimensionality of the gas.  

To determine the relative contribution of density changes to the non-Gaussian
velocity distributions in the gas, the number of particles on the plate was
increased by $15\%$ and decreased by $10\%$ from the value of 
$\rho = 0.532$ and the lid was kept on.  For all accelerations, the 
flatness decreased with increased density.  This surprising result may be
related to the fact that strongly non-Gaussian 
distributions observed at low $\Gamma$ are accompanied by strong clustering 
\cite{us}.  If the average density is increased, the larger excluded volume
means that less phase space remains for fluctuations to persist.  The fact
that increasing the density with the lid on makes the velocity distribution
more Gaussian suggests that the crossover to Gaussian observed without the
lid is not due to the decrease in density of the gas.

Puglisi {\em et al.} \cite{puglisi} have proposed a model which relates strong 
clustering to non-Gaussian velocity distributions in a driven granular
medium.  In their framework, at each local density the velocity distributions 
are Gaussian, and the non-Gaussian behavior arises from the relative weighting
of the temperature by local density in the following manner:

\begin{equation}
P(v) = \sum_{boxes} n(N) e^{-(\frac{v^2}{v_0^2(N)})}
\label{eq:density}
\end{equation}
\noindent
where $v_0^2(N)$ is the second moment of the distribution for the number of  
boxes, {\em n}, that contain N particles.  In this model, the local temperature
is a decreasing function of the local density, and the velocity distributions
conditioned on the local density are Gaussian. 

In our experiment, this feature can be examined by conditioning
the local horizontal granular temperature on the local density.  That is,
examining the distribution of velocities for data at a constant
number of particles in the frame of the camera in the strongly 
clustering regime at $\Gamma \approx 0.8$ \cite{us}.  
In the strongly clustering regime, we do observe a direct correlation
between local density and temperature.  Figure \ref{fig:four} 
is a plot of the local temperature as a function of particle number in the
camera frame normalized by the granular temperature, including data from the 
open system for both the smaller and larger particles as well as for the 
confined system using the larger particles.  

The result is similar to the model of Puglisi {\em et al.} 
\cite{puglisi}:  At low T$_G$, when the particle-particle
correlations are strongest (and larger than those of an equilibrium
hard sphere gas \cite{us}), there is a density dependence to the granular
temperature (filled circles).  At $\Gamma = 3$, where all of the particles are 
essentially uncorrelated in a 3D volume in the absence of a lid, there 
is no density dependence (open circles, stars).  However, even in the confined 
system at $\Gamma = 3$, where  the distribution is still not Gaussian, no 
appreciable density dependence is observed (diamonds), suggesting that the
non-Gaussian velocity distributions and density-dependent temperature are
not as simply dependent upon one another as they are in the model of 
Puglisi {\em et al.} \cite{puglisi}.  
In fact, while there is a clear density dependence on the local temperature
at low $\Gamma$, the measured velocity distribution conditioned on the local 
temperature is not Gaussian.
At each density, the velocity distribution function is almost identical to 
that of the whole:  when the entire distribution is non-Gaussian, the 
distribution at a single density is non-Gaussian, and when the whole is 
Gaussian, each conditional velocity distribution is Gaussian.  

A more general form of Eq. \ref{eq:density} represents the total velocity 
distribution as a product of local Gaussian velocity distributions with
a distribution of local temperatures:

\begin{equation}
P(v) = \int_{\vec{r}(t)} f(T(\vec{r}(t))) e^{-(\frac{v^2}{T(\vec{r}(t))})} d\vec{r}dt
\label{eq:tempdens}
\end{equation}
\noindent
where $T(\vec{r}(t))$ is the local temperature that is varying in space and
time.  Conditioning on the local temperature would then recover the 
underlying Maxwell statistics in the fluctuations \cite{BizonSSS99}.  

Performing this analysis on our data does not succeed in producing Gaussian
statistics.  Within small windows of local temperature, the distributions 
remain non-Gaussian.  Indeed, the analysis can be extended to condition on
both the local temperature and density in the system, but with similarly 
limited success except for the slowest of particles in the most dilute regions 
of the system, although all of the conditioned distributions are closer to 
Gaussian than the full distribution.
Curiously, if the local velocities are normalized by the magnitude of the 
local granular temperature, then the total distribution is Gaussian 
\cite{BizonSSS99}.

These results demonstrate several important new characteristics of a 2D 
granular gas.  Extreme non-Gaussian velocity distributions observed previously 
\cite{us} involve strong cross-correlations between density and velocity
fluctuations.  This is not the whole picture, however, as a 2D constrained
granular gas also demonstrates strongly non-Gaussian behavior without 
a significant cross-correlation between density and velocity distributions.
Finally, the crossover to Gaussian distributions observed when the gas
is fully 3D is probably associated with the change in dimensionality or
momentum transfer, rather than the increase in kinetic energy or the decrease
in density.  

This work was supported by an award from the Research Corporation, a grant 
from the Petroleum Research Fund and grant DMR-9875529 from the
NSF.  One of us (JSU) was supported by a fellowship from the Sloan Foundation.

%
\begin{figure}
\caption{Log-linear plot of velocity distribution functions for increasing 
$\Gamma$ at constant frequency.  
As the acceleration is increased, the distributions go from having 
nearly exponential to Gaussian tails.
($\bigcirc$) $\Gamma = 0.93$,($+$) $\Gamma = 1.5$, ($\Box$) 
$\Gamma = 3.0$.
}
\label{fig:one}
\end{figure}


\begin{figure}
\caption{(a) Pair correlation functions for the velocity distributions in 
Figure \ref{fig:one} ($\bigcirc$) $\Gamma = 0.93$,
($+$) $\Gamma = 1.5$, ($\triangle$) $\Gamma = 2.0$, ($\Box$) $\Gamma = 3.0$. 
(b) Pair correlation functions for the velocity distributions where a lid
constrains the system to remain 2D.  The particle 
correlations remain as the shaking amplitude is increased.
}
\label{fig:two} 
\end{figure}


\begin{figure}
\caption{(a) Plot of the horizontal granular temperature, T$_G$, as a function 
of dimensionless acceleration, $\Gamma$, with and without a lid.
Data for $\rho = 0.532$: ($\bigcirc$) d = 1.2 mm and ($\ast$) d = 1.6
mm without lid; ($\Diamond$) d = 1.6 mm with lid.  Data for $\rho = 0.478$ 
($\triangle$) and $\rho = 0.611$ ($\Box$) with a lid and d = 1.6 mm.   
(b) Plot of the flatness as a function of granular temperature with
and without a lid in the system.  At low $\Gamma$ the system is nearly 2D 
and the lid has no effect.
}
\label{fig:three} 
\end{figure}


\begin{figure}
\caption{Plot of $T(N)/T_G$ where N is the number of particles in the 
camera frame (see text).  At low $\Gamma$, the local temperature and density 
are strongly correlated.  
At high $\Gamma$, the local granular temperature is independent 
of density, even when the velocity distributions remain non-Gaussian in the 
constrained system.
}
\label{fig:four}
\end{figure}


\end{document}